\documentclass[twocolumn,showpacs,preprintnumbers,amsmath,amssymb,pre]{revtex4}

\usepackage{textcomp}
\usepackage{graphicx}
\usepackage{dcolumn}
\usepackage{bm}
\usepackage{psfrag}

\begin{document}

\preprint{APS/123-QED}

\title{Simulations of the kinematic dynamo onset of spherical Couette flows with
smooth and rough boundaries}

\author{K. Finke}

\author{A. Tilgner}

\affiliation{Institute of Geophysics, University of G\"ottingen,
Friedrich-Hund-Platz 1, 37077 G\"ottingen, Germany }

\date{\today}

\begin{abstract} We study numerically the dynamo transition of an
incompressible electrically conducting fluid filling the gap between
two concentric spheres. In a first series of simulations, the fluid
is driven by the rotation of a smooth inner sphere through no-slip
boundary conditions, whereas the outer sphere is stationary. In a second series
a volume force intended to simulate a rough surface drives the fluid
next to the inner sphere within a layer of thickness one tenth of the
gap width. We investigate the
effect of the boundary layer thickness on the dynamo
threshold in the turbulent regime. The simulations show that the
boundary forcing simulating the rough surface lowers the necessary rotation rate, which
may help to improve spherical dynamo experiments.
\end{abstract}

\maketitle

\section{Introduction}

Dynamo theory describes the generation of magnetic fields in 
flows of conducting fluids, for example in stars or planetary interiors. Several
experiments have been built in order to reproduce a dynamo in the
laboratory. After the first experiments in Riga \cite{Gailit00} and Karlsruhe
\cite{Muller02}, more recent
experiments \cite{Moncha07,Lathro11} implement flows less confined to specific
shapes than the first experiments, so that the effect of
turbulence on dynamo action is more relevant. 
Nevertheless, the main contribution to magnetic field generation in these
experiments is still assumed to come from the time averaged part of the flows. The numerical
investigations presented in this paper are directly applicable to spherical
experiments like those in Maryland \cite{Lathro11}.

The spherical Couette flow with a stationary outer sphere is mainly a
differential rotation driven by the inner sphere, together with a jet in the
equatorial plane in which fluid is centrifuged radially outward. A flow
along the rotation axis connecting the poles of the two spheres brings the
fluid back to the inner sphere. The flow thus consists of differential rotation
and two loops in the meridional circulation. It is helical with opposite
helicities in the two hemispheres and is topologically equivalent to the s2t1
flows studied by Dudley and James \cite{Dudley89} which are known to be dynamos. The
spherical Couette flow on the contrary does not generate magnetic fields for
Reynolds numbers $\mathrm{Re}$ up to a critical value where the flow becomes unstable
\cite{Guervi10}. It
therefore seems possible that turbulent spherical Couette flow is a small scale
dynamo. In this type of dynamos, at magnetic Prandtl numbers $\mathrm{Pm}$ with
$\mathrm{Pm} \gg 1$, the largest production of the magnetic field
occurs at the resistive scale and the critical magnetic Reynolds number $\mathrm{Rm}$ increases
with increasing hydrodynamic Reynolds number. For $\mathrm{Pm} \ll 1$,
the largest production of magnetic field occurs at some scale larger than 
the viscous scale \cite{Scheko04,Iskako07},
so that the critical $\mathrm{Rm}$ is independent of $\mathrm{Re}$ for
$\mathrm{Re}$ large enough, because increasing $\mathrm{Re}$ then adds small
vortices with magnetic Reynolds numbers less than 1. The magnetic field shows
nothing but diffusive dynamics on these length scales so that they do not
contribute to either creation or mixing of magnetic field. Ref. \onlinecite{Guervi10}
found $\mathrm{Rm_c}$ to increase with $\mathrm{Re}$ for the entire investigated
parameter range, which further adds to the suspicion that spherical Couette
flows could be small scale dynamos. Since the Reynolds number
$\mathrm{Re}$ in numerical simulations is limited, the magnetic Prandtl number
$\mathrm{Pm}$ has to be adjusted to values far above the $\mathrm{Pm}$ of liquid
sodium (which is the liquid commonly used in experiments) in order to achieve
high enough magnetic Reynolds numbers. There is always a need to
extrapolate from numerical results to larger $\mathrm{Re}$, so that it is always
important to understand the dependence of the dynamo threshold on $\mathrm{Re}$
and $\mathrm{Pm}$ \cite{Scheko04,Iskako07,Ponty05}.

An experiment is under construction which realizes spherical Couette flow in
liquid metals \cite{Triana11}. One motivation for the present study is to predict whether this
experiment will be able to sustain a self-generated magnetic field. This
question will be investigated in section \ref{no_slip}. Because of the
pessimistic answer obtained in this section, section \ref{rough} considers the possible improvement
obtained by welding blades on the inner sphere in order to strengthen the
coupling between the fluid and the rotation of the inner sphere.

\section{The mathematical model \label{Math}}

The system under investigation consists of
two concentric speres with radii $R_i$ and $R_o$ forming a gap of width
$R_i - R_o = d$ and aspect
ratio $R_i/R_o = 1/3$. The fluid in the gap is driven by the rotation of the
inner core, which
rotates at angular frequency $\Omega_i$ about the $z-$axis, while the
outer boundary is at rest. The fluid is characterized by its kinematic viscosity
$\nu$ and its magnetic diffusivity $\lambda$. The
evolution of the magnetic field in an incompressible electrically
conducting fluid is described by the induction equation. The system is governed
by two numbers, the Reynolds number $\mathrm{Re}$, and the magnetic Prandtl
number $\mathrm{Pm}$ or alternatively the magnetic Reynolds number $\mathrm{Rm}$
defined by
\begin{equation}
\mathrm{Re} = \frac{\Omega_i d^2}{\nu}, \qquad \mathrm{Pm} =
\frac{\nu}{\lambda}, \qquad \mathrm{Rm} = \mathrm{Re} \text{ }
\mathrm{Pm}.
\end{equation}
From now on, only adimensional variables will be used in this paper.
Choosing as units of
time and length the reciprocal of the inner core's rotation rate and the gap
width, the induction equation reads for non-dimensional magnetic and velocity fields
$\bm{B}$ and $\bm{v}$:

\begin{equation} 
\partial_t \bm{B} + \nabla \times (\bm{B} \times \bm{v}) =
\frac{1}{\mathrm{Re}\text{ }\mathrm{Pm}} \nabla^2 \bm{B} ~~~,~~~
\nabla \cdot \bm{B} = 0. 
\label{eq:Induction}
\end{equation}

The velocity field itself is a solenoidal vector field
determined by the Navier-Stokes equation and the continuity equation:

\begin{equation} 
\partial_t \bm{v} + (\bm{v} \cdot \nabla ) \bm{v} = - \nabla
\Phi + \frac{1}{\mathrm{Re}} \nabla^2 \bm{v} \text{ }+ \bm{F} 
~~~,~~~
\nabla \cdot \bm{v} = 0 .  
\label{eq:Navier_Stokes} 
\end{equation}

Eq. (\ref{eq:Navier_Stokes}) contains a volume force $\bm F$ and $\Phi$ stands
for the pressure variable.
Eqs. (\ref{eq:Induction}) and (\ref{eq:Navier_Stokes}) describe the kinematic
dynamo problem, in which one assumes the system to be near the onset of magnetic
field generation, so that the magnetic field strength is small and the Lorentz
force is negligible in equation (\ref{eq:Navier_Stokes}). 

The parameters $\mathrm{Re}$ and $\mathrm{Rm}$ depend on the inner core's
rotation rate. More revealing parameters are the
Reynolds numbers $\mathrm{\overline{Re}}$ and $\mathrm{\overline{Rm}}$
based on the rms velocity $\overline{v_{\text{rms}}}$ defined as
\begin{equation}
\overline{v_{\text{rms}}}=\sqrt{2 E_{\mathrm{kin}}/V}
~~~,~~~
E_{\mathrm{kin}}=\langle \int \frac{1}{2} \bm v^2 dV \rangle
\end{equation}
and 
\begin{equation}
\mathrm{\overline{Re}} = \mathrm{Re} \text{ } \overline{v_{\text{rms}}}
~~~,~~~
\mathrm{\overline{Rm}} = \mathrm{\overline{Re}} \text{ }
\mathrm{Pm}
\end{equation}
where $V$ is the volume of the shell, brackets denote time average and
$E_{\mathrm{kin}}$ the kinetic energy.

In the following the effect of different surface properties on the dynamo
threshold are compared. The inner and outer spheres have radii $r_i=1/2$ and
$r_o=3/2$, respectively. The inner sphere and the
space surrounding the outer sphere are assumed to be insulating. Two different types
of boundary forcing are
implemented in order to simulate both smooth and rough surfaces. For the
smooth surface no-slip boundary conditions 
\begin{equation}
\bm v = \hat{\bm z} \times \bm r ~~~ \mbox{at} ~~~ r=r_i
~~~ , ~~~
\bm v = 0 ~~~ \mbox{at} ~~~ r=r_o
\end{equation}
are chosen together with $\bm F=0$ in eq. (\ref{eq:Navier_Stokes}).
In the second case, the inner boundary is assumed free slip and the
fluid is driven in one tenth of the gap width by a toroidal force field
$\bm F$ of spherical harmonic degree $l=1$ and order $m=0$, given 
in spherical polar coordinates $(r,\theta,\varphi)$ by:

\begin{equation}
\bm{F} = \frac{1}{2} \left[ 1-\tanh \left(\frac{60}{d}  (r - r_i - \frac{d}{10})
\right) \right] \sin \theta  \text{ }
\hat{e}_\varphi.
\end{equation}

This forcing qualitatively
reproduces the flow driven by an inner sphere with blades 
of height one tenth the gap size mounted along meridians. The non-dimensional
rotation rate of the inner sphere, $\Omega'_i$, has to be found a posteriori
in these simulations by computing

\begin{equation}
\Omega'_i= \frac{3}{8 \pi r_i} \int_0^{2\pi} d \varphi \int_0^\pi  d \theta~ \sin^2 \theta
\langle v_{\varphi}(r=r_i,\theta,\varphi,t) \rangle
\end{equation}
with which a Reynolds number $\mathrm{Re'}$, analogous to
$\mathrm{Re}$ for the no slip boundaries, is defined 
as
\begin{equation}
\mathrm{Re'}= \mathrm{Re} \text{ } \Omega'_i
\end{equation}

In the following $\mathrm{Re}$ is going to be increased up to $1.67
\cdot 10^{4}$ for the no-slip boundary conditions and up to $2.5 \cdot 10^{3}$
in the simulations utilizing the volume force.
The magnetic Prandtl number is of order unity for the simulations just at
the onset of magnetic amplification. The
numerical method is a spectral method, in which the fields are expanded in
spherical harmonics and Chebychev polynomials \cite{Tilgne99}.

\section{No-slip \label{no_slip}}

\subsection{Hydrodynamic Characteristics}

\begin{figure}[h] \centering
\includegraphics[width=0.65\linewidth]{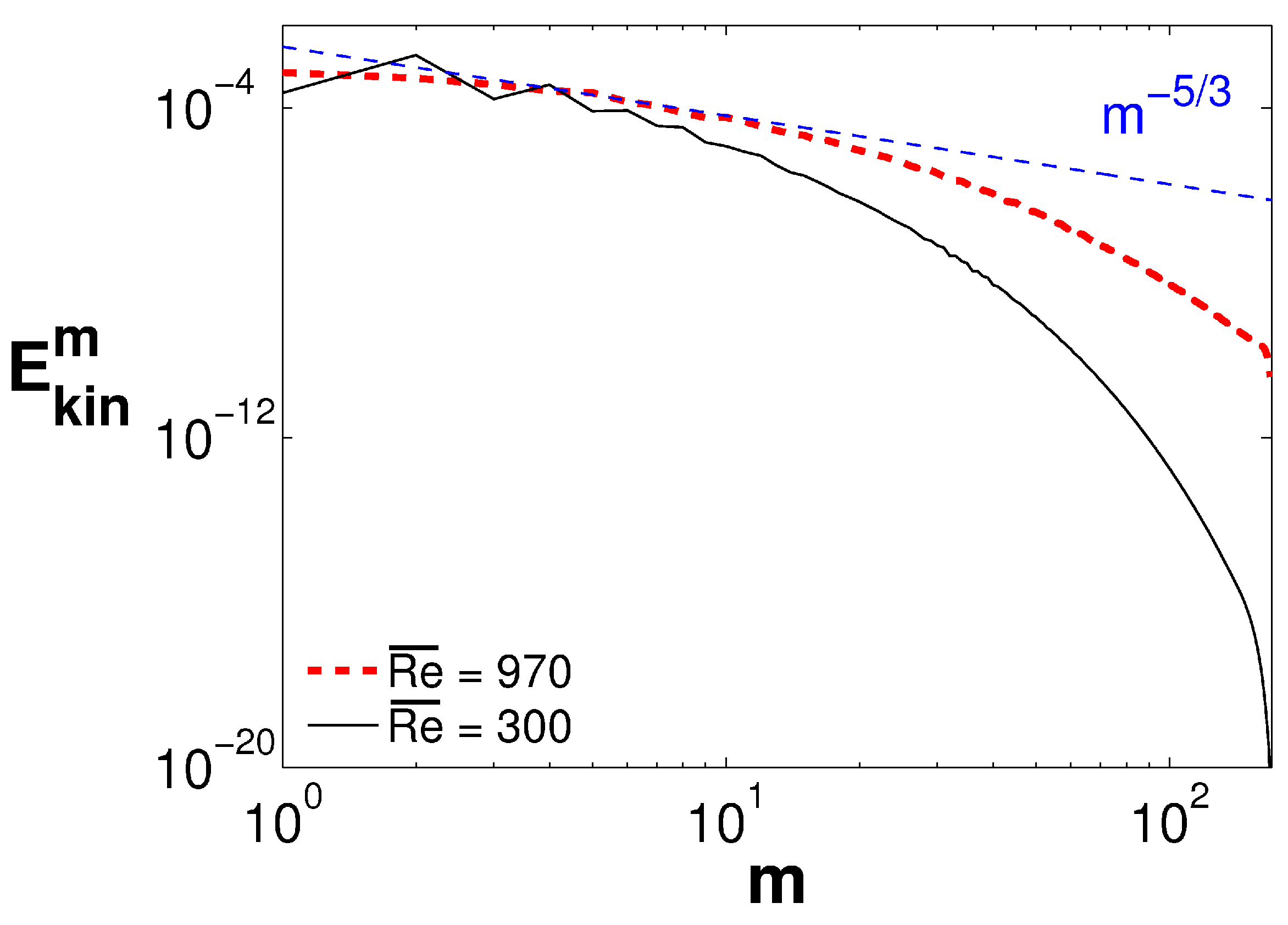}
\hfill
\includegraphics[width=0.65\linewidth]{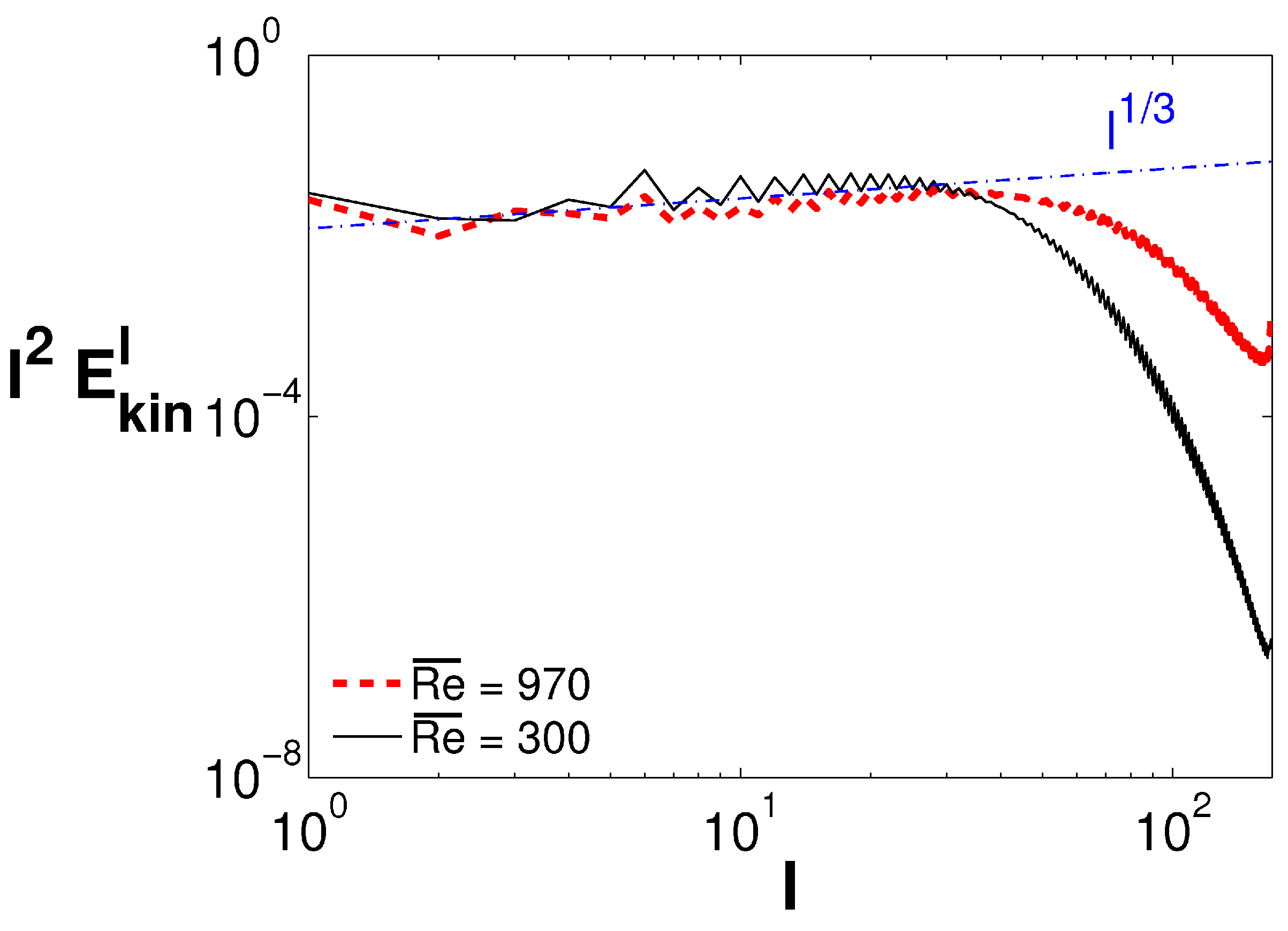}
\caption{(Color online) Power
spectrum of kinetic energy plotted against spherical harmonic degree $m$ (top) and the turbulent 
rate of strain versus spherical harmonic order $l$ (bottom)
for $\mathrm{\overline{Re}} = 300$ (black continuous line) and $970$ (thick red dashed line),
together with the Kolmogorov power laws $m^{-5/3}$ and $l^{1/3}$ (blue dashed line).}
\label{fig:spectra_compare_E} 
\end{figure}

The hydrodynamic properties of the spherical Couette flow with stationary outer
sphere have already been investigated in detail in
\cite{Dumas94,Holler06,Guervi10}. At low Reynolds numbers the basic spherical
Couette flow is axisymmetric and we find a critical Reynolds number of $\mathrm{Re_h}
= 1500$ ($\mathrm{\overline{Re}_h} = 105$) beyond which 
small non-axisymmetric perturbations
increase and an instability develops as a propagating wave on the equatorial jet
with a dominant azimuthal wavenumber $m=2$, which is in agreement with
\cite{Guervi10,Holler06}. At $\mathrm{Re_s} = 2800$ ($\mathrm{\overline{Re}_s} =
178$) a second transition occurs. Beyond this value, the amplitudes of odd
wavenumbers $m$ of the kinetic energy develop as well and the
power spectrum begins to flatten and approaches power laws in
$m^{-5/3}$ and $l^{-5/3}$, indicative of Kolmogorov turbulence (see Figure \ref{fig:spectra_compare_E}).
Large scales of the velocity field, however, have comparable spectra for all
$\mathrm{Re}$. The bottom panel shows the power spectra of an estimate of the 
turbulent rate of strain of spherical harmonic 
order l, which is simply the kinetic energy multiplied by $l^2$ and has its maximum at the viscous scale.
At $\mathrm{\overline{Re}} = 970$ the inertial range reaches up to $l \approx 40$.

\begin{figure}[h] \centering
\includegraphics[width=0.65\linewidth]{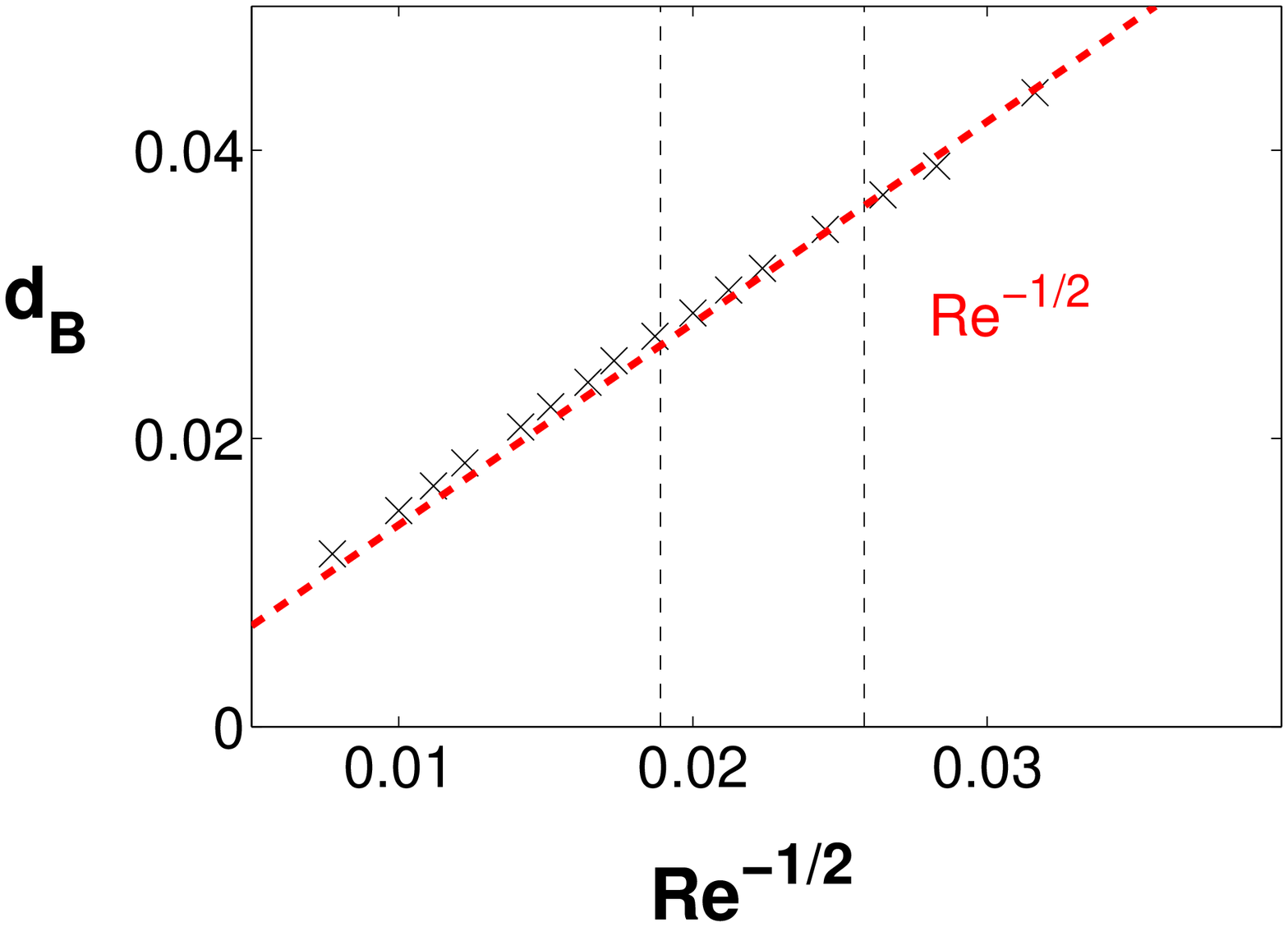}
\hfill
\includegraphics[width=0.65\linewidth]{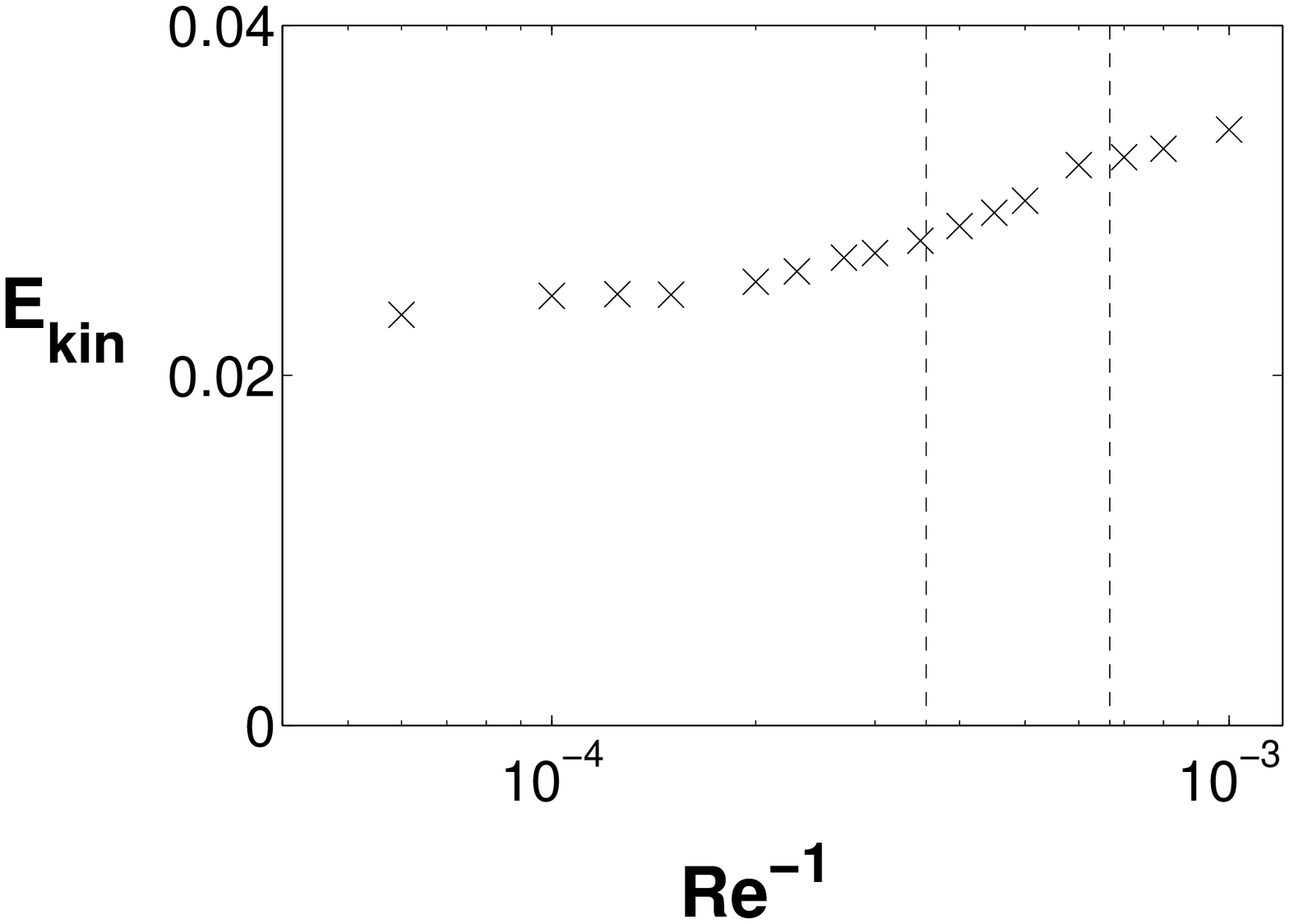} 
\hfill
\includegraphics[width=0.65\linewidth]{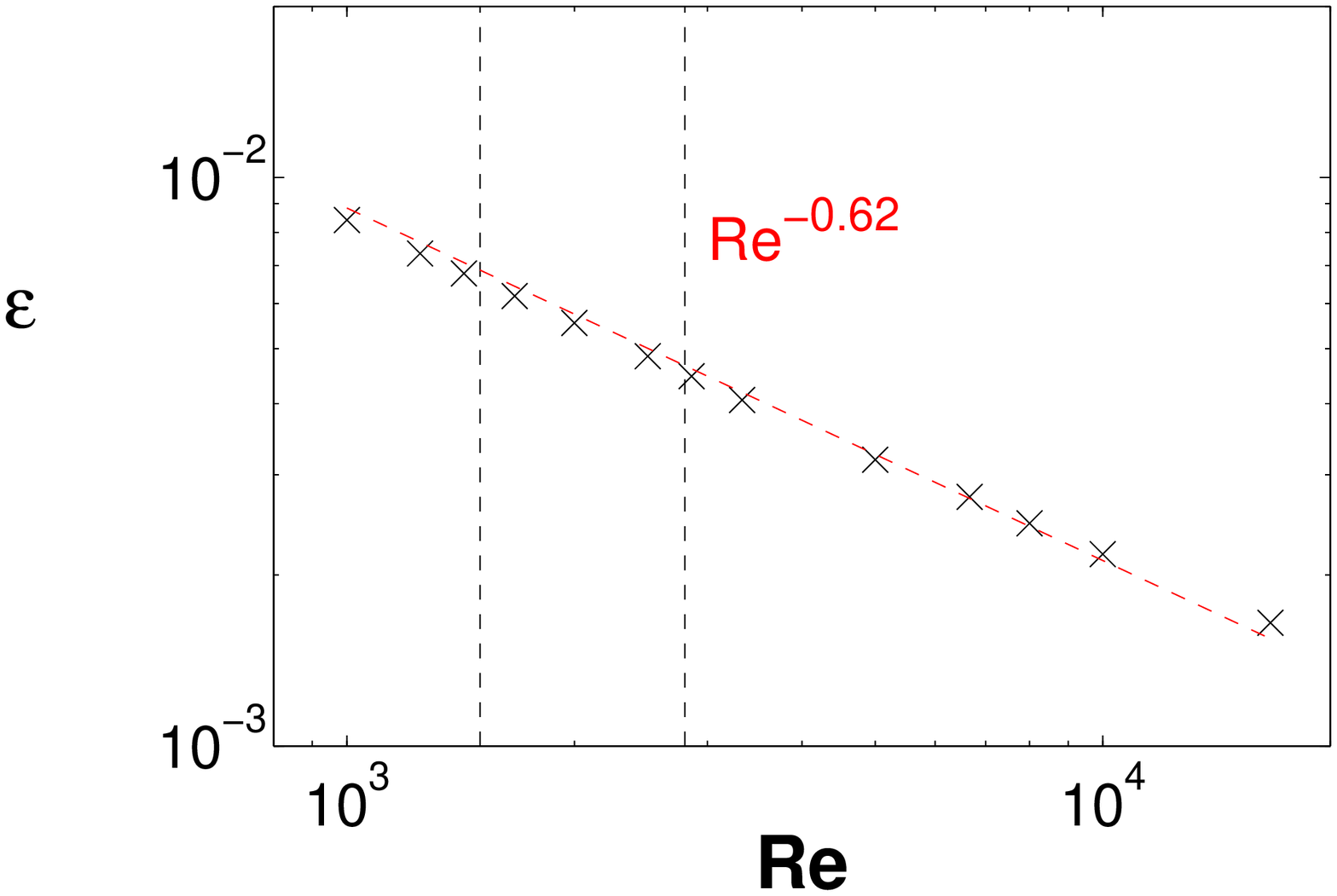} 
\caption{(Color online) Boundary layer thickness $d_B$ plotted against
$\mathrm{Re}^{-1/2}$ (top), kinetic
energy plotted against $\mathrm{Re}^{-1}$ (middle), and energy dissipation rate plotted against $\mathrm{Re}$ (bottom). The
vertical dashed lines mark $\mathrm{Re_h}$ and $\mathrm{Re_s}$.} 
\label{fig:blthickness_ns} 
\end{figure}

The wavenumber of the most unstable mode depends on the aspect ratio of the
shell and switches from $m=2$ to $m=3$ at an aspect ratio close to the one
chosen here \cite{Holler06}. It is possible to generate flows with $m=3$ as dominating mode by
starting from an equilibrated solution at $\mathrm{Re}>\mathrm{Re_s}$ and
lowering $\mathrm{Re}$ to some value in between $\mathrm{Re_h}$ and
$\mathrm{Re_s}$.

Figure \ref{fig:blthickness_ns} (top) shows the
thickness of the boundary layer of $u_\varphi$ near the inner core. In order to
produce this figure, $u_\varphi$ has been averaged over spherical surfaces and the
boundary layer thickness defined as the distance from the inner sphere at which
this averaged velocity drops to the arithmetic mean of its value at the boundary and its
value averaged over radius. At high $\mathrm{Re}$ the evolution of the boundary
layer thickness is approaching $\mathrm{Re}^{-1/2}$, as one would expect from
theory.

Figure \ref{fig:blthickness_ns} (middle) shows the dependence of the dimensionless
kinetic energy on $\mathrm{Re}$, which converges to a constant in the limit of
$\mathrm{Re}^{-1} \rightarrow 0$. The viscosity becomes irrelevant for high
$\mathrm{Re}$ and the only remaining control parameters entering the
dimensional kinetic energy
are the density and $\mathrm{\Omega_i}$, so that the kinetic energy scales with
$\Omega_i^2$ and the dimensionless kinetic energy becomes constant.

A quantity of direct relevance to experiments is the energy dissipated in the
flow. The energy budget, obtained by taking the scalar product of eq.
(\ref{eq:Navier_Stokes}) with $\bm v$ and integrating over space, reads

\begin{equation}
\partial_t \int \frac{1}{2} \bm v^2 dV = \frac{1}{\mathrm{Re}} \tau
-\frac{1}{\mathrm{Re}} \int (\partial_i v_j)^2 dV + 
\int \bm F \cdot \bm v dV
\label{eq:energy_budget}
\end{equation}
with
\begin{equation}
\tau = - \int_0^{2\pi} d \varphi \int_0^\pi d \theta r^3 \sin^2 \theta
\left(\partial_r v_\varphi - \frac{v_\varphi}{r} \right),
\label{equ:torque}
\end{equation}
evaluated on the inner boundary $r=r_i$,
being the torque on the inner boundary. $\bm F =0$ for the simulations in this
section so that the time averaged torque $\langle \tau \rangle$ is directly
related to the energy dissipation $\epsilon$ by
$\epsilon = \langle \tau \rangle/\mathrm{Re}$.
As shown in fig. \ref{fig:blthickness_ns},
the dissipation scales as $\mathrm{Re}^{-0.62}$. The bracket in the integrand in
eq. (\ref{equ:torque}) is dominated by the derivative $\partial_r v_\varphi$
which can be estimated as $v_{\mathrm{rel}}/d_B$, where $d_B$ is the boundary
layer thickness and $v_{\mathrm{rel}}$ the velocity of the inner boundary
relative to the bulk fluid. Since $d_B \propto \mathrm{Re}^{-1/2}$, we have
$\epsilon \propto \mathrm{Re}^{-1/2}
v_{\mathrm{rel}}$. If the motion of the inner core was completely decoupled from
the rotation of the inner boundary, $v_{\mathrm{rel}}$ would be independent of
$\mathrm{Re}$ and $\epsilon \propto
\mathrm{Re}^{-1/2}$. There is of course some entrainment of the fluid by the
rotation rate of the inner sphere and one finds an exponent of -0.62 instead of
-0.5.

\subsection{Dynamo transition}

\begin{figure}[h] \centering
\includegraphics[width=0.95\linewidth]{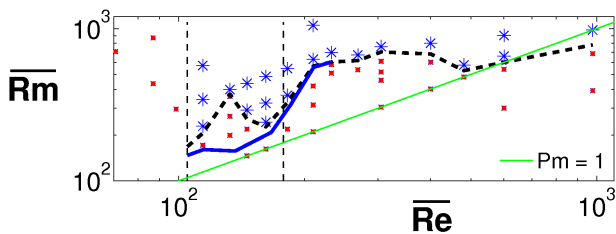}
\caption{(Color online) Dynamo transition for magnetic fields dominated by modes with
$m=2$ (black dashed line) and $m=3$ (blue bold line).
Failed dynamos are indicated by red dots, working dynamos by blue asterisks.
The vertical dashed lines
seperate the different hydrodynamic regimes: Axissymmetric flow (left), first nonaxisymmetric
instability (middle) and turbulent regime (right).} \label{fig:Dynamo_onset_ns}
\end{figure}

The dynamo transition for the spherical Couette flow with stationary outer
sphere has already been computed in \cite{Guervi10} with different
magnetic boundary conditions and an aspect ratio of $r_i/r_o= 0.35$. These results
will be compared with ours in the last section. Figure \ref{fig:Dynamo_onset_ns} shows
the dynamo simulations in the $(\mathrm{\overline{Re}},\mathrm{\overline{Rm}})-$plane.
Asterisks mark working dynamos and dots are failed
dynamos. Linear interpolation between the growth rates computed at those points allows
us to find the locus of zero growth rate. The onset of magnetic field
amplification obtained in this way is indicated by the 
dashed line. The magnetic fields are dominated by modes with an azimuthal
wavenumber of $m=2$. As mentioned above, suitable initial conditions lead to
magnetic fields dominated by modes with 
$m=3$. The onset for these dynamos is indicated by the thick solid line.
The two lines coincide 
for $\mathrm{\overline{Re}} > \mathrm{\overline{Re}_s}$. The straight
line indicates $\mathrm{Pm} = 1$. Except for the highest $\mathrm{\overline{Re}}$ the
magnetic Prandtl number is always larger than one. In agreement with
\cite{Guervi10} there are no working dynamos for axisymmetric flows. For
$\mathrm{\overline{Re}_h} < \mathrm{\overline{Re}} < \mathrm{\overline{Re}_s}$ the critical
magnetic Reynolds number $\mathrm{\overline{Rm}_c}$ for $m=2$ first increases,
reaches 
a maximum and finally decreases, until at the second transition odd modes of
the kinetic energy become unstable and the dynamo threshold again increases.

The non-monotonous dependence of
$\mathrm{\overline{Rm}}$ on $\mathrm{\overline{Re}}$ in the interval 
$\mathrm{\overline{Re}_h} < \mathrm{\overline{Re}} < \mathrm{\overline{Re}_s}$ appears to
be related to two other features of the dynamo: The drift frequency of the
unstable mode and the dominating wavenumber in the magnetic field.

The velocity field for $\mathrm{\overline{Re}_h} < \mathrm{\overline{Re}} <
\mathrm{\overline{Re}_s}$ is dominated by one azimuthal wavenumber and its harmonics.
Time series of the radial velocity field at a fixed point show that the phase
velocity of the propagting wave diminishes for increasing $\mathrm{\overline{Re}}$.
Similarly to previous studies of kinematic dynamos \cite{Tilgne08,Reuter09}, it seemed useful to run
simulations in which the velocity field is a snapshot taken from the full
simulation, and which is set to drift at arbitrary phase velocities in order
to investigate the dependence of the growth rate on the phase
velocity of the instability. It turned out that the critical magnetic Reynolds 
number increases with the phase velocity, as does $\mathrm{Rm_c}$ in Figure
\ref{fig:Dynamo_onset_ns}. The time evolution of the magnetic energy using the  
artificially drifting velocity field shows the same characteristic superposition
of an exponential growth and an oscillation as in the dynamic simulations in that
region of $\mathrm{\overline{Re}}$, but neither the frequency of the oscillation
of the magnetic energy nor the growth rate agree exactly.
However, the amplitude of the $m=4$ contributions to the kinetic
energy are only one order of magnitude smaller than the dominant mode and they
introduce a slight time dependence in the codrifting frame of reference,
so that the simulations of the drifting snapshot can only catch the
qualitative features of the full simulation.

\begin{figure}[h] \centering
\includegraphics[width=0.65\linewidth]{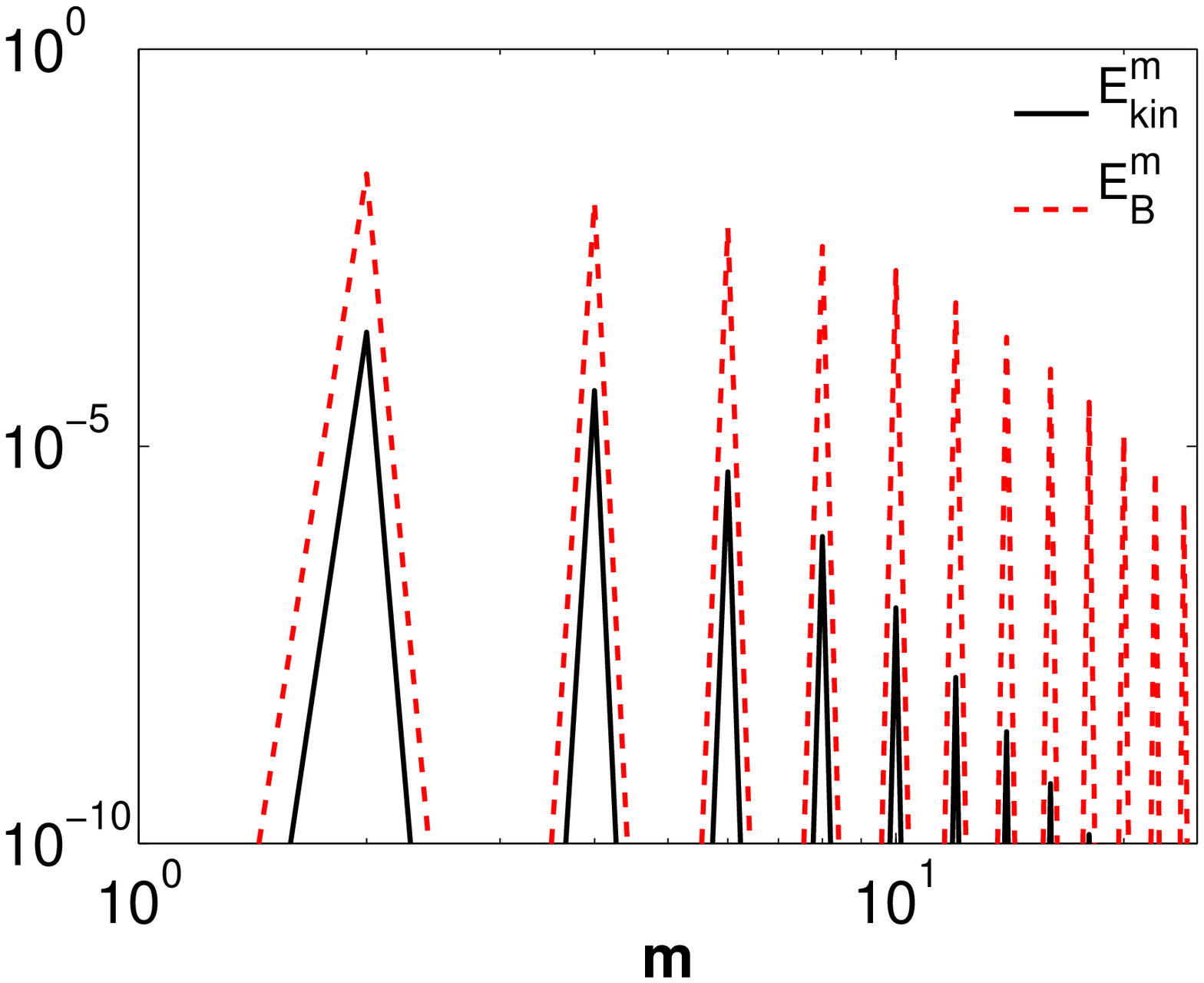}
\hfill
\includegraphics[width=0.65\linewidth]{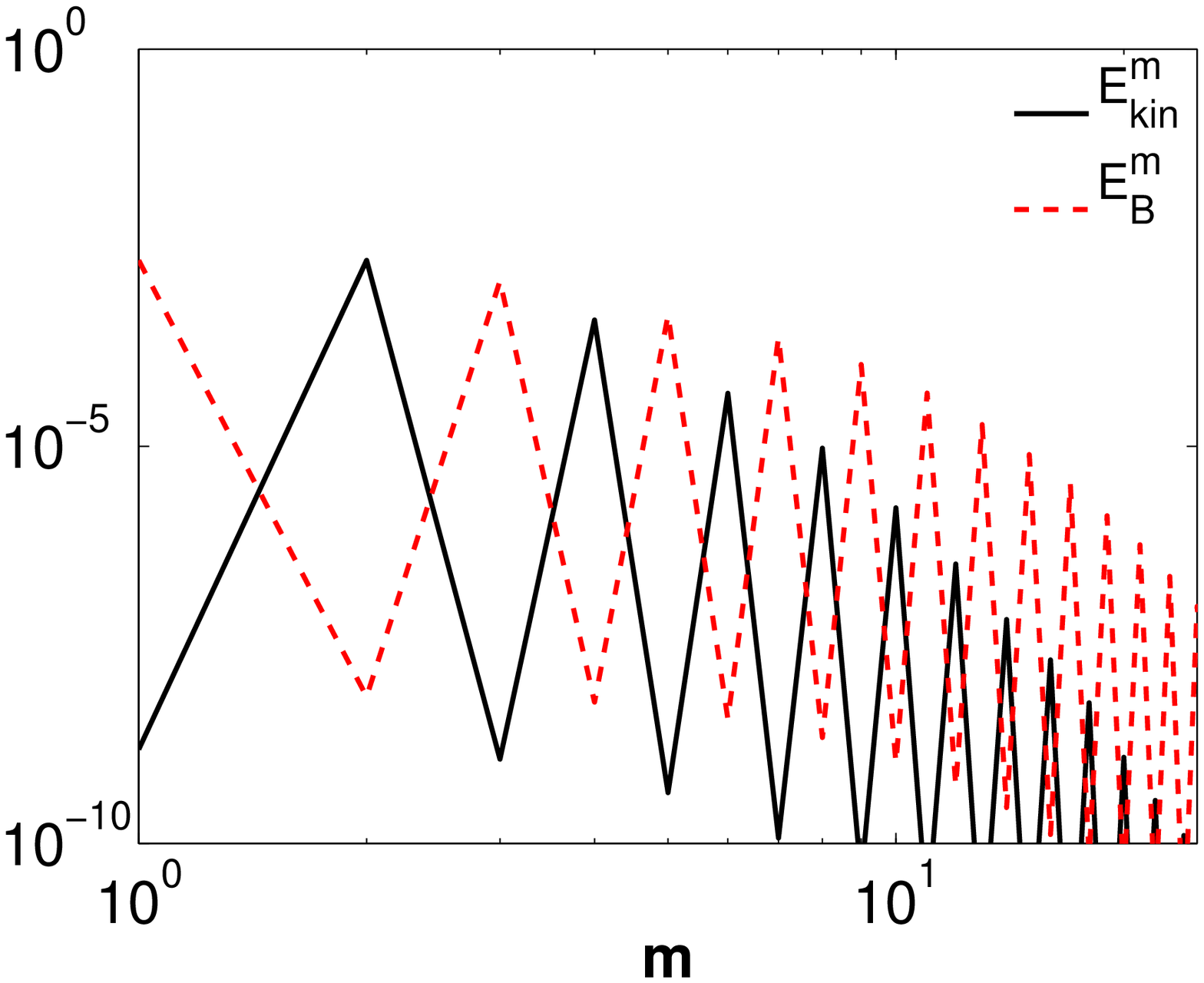} 
\caption{(Color online) Power
spectrum of the kinetic energy (black continuous line) and the magnetic energy (red dashed line)
for $\mathrm{\overline{Re}} = 130$ (top) and $\mathrm{\overline{Re}} = 160$ (bottom)}
\label{fig:spec_even} \end{figure}

\begin{figure}[h] \centering
\includegraphics[width=0.7\linewidth]{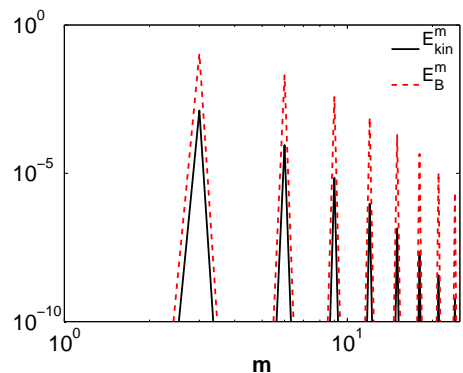} 
\caption{(Color online) Power
spetrum of the kinetic energy (black solid line) and the magnetic energy (red dashed line)
for $\mathrm{\overline{Re}} = 130$ in a simulation in which modes with $m=3$
dominate.} \label{fig:spec_third} \end{figure}

In Figure \ref{fig:spec_even} power spectra of the kinetic and magnetic energy of 
the simulations at $\mathrm{\overline{Re}} = 130$ and $\mathrm{\overline{Re}} = 160$
are shown. In the first case only even wavenumbers of the magnetic energy 
are amplified, whereas odd wavenumbers decreases in time. The opposite is seen in 
the second case, where odd wavenumbers of the magnetic energy are amplified and even 
wavenumbers decrease. The different symmetry of the magnetic field seems to be 
responsible for the decrease of the dynamo threshold, since this decrease correlates 
with the appearence of the odd wavenumbers in the magnetic field. In addition, Figure
\ref{fig:spec_third} shows the spectra of the simulations with $m=3$ as the dominant wavenumber,
which have for $\mathrm{\overline{Re}_h} < \mathrm{\overline{Re}} <
\mathrm{\overline{Re}_s}$ the same symmetry and accordingly the dynamo threshold is monotonously
increasing.

For $\mathrm{\overline{Re}}>\mathrm{\overline{Re_s}}$, the critical $\mathrm{\overline{Rm}}$
first increases but then reaches a plateau at $\mathrm{\overline{Rm}}=600$. $Pm$ is
less than 3 throughout the plateau region. It will now be argued that even more
turbulent flows will not alter the critical $\mathrm{\overline{Rm}}$ significantly.

\begin{figure}[h] \centering
\includegraphics[width=0.65\linewidth]{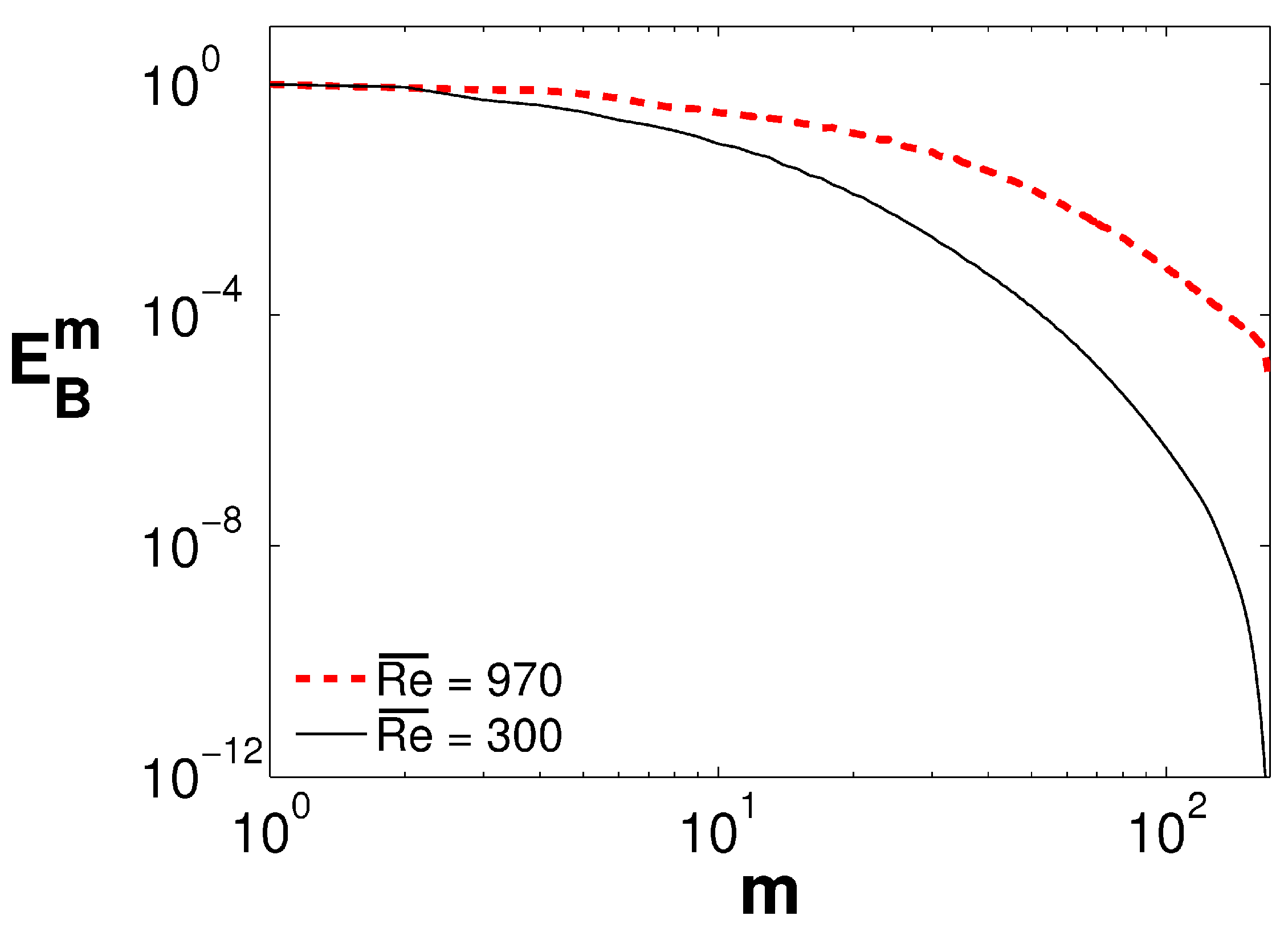}
\hfill
\includegraphics[width=0.65\linewidth]{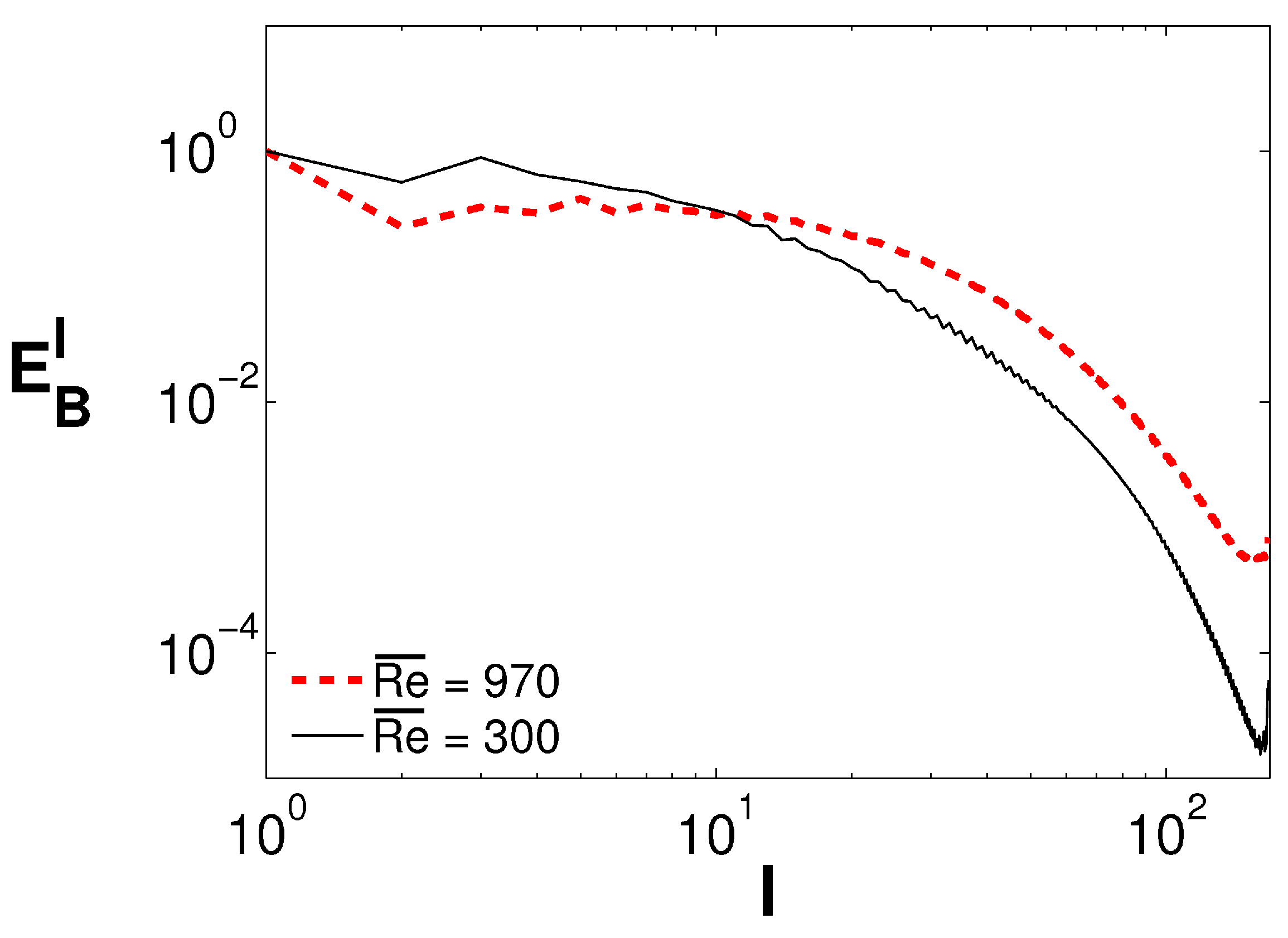}
\caption{(Color online) Power
spectrum of magnetic energy of spherical harmonic degree $m$ (top) and order $l$ (bottom) 
for $\mathrm{\overline{Re}} = 300$ (black continuous line) and $970$ (red dashed line).}
\label{fig:spectra_compare_B} 
\end{figure}

In the plateau region, the Reynolds number is high enough for the kinetic energy
spectrum to decay as a power law as a function of both $m$ and $l$ with an
exponent of -5/3. The phenomenolgy developed for small scale dynamos in
Kolmogorov turbulence \cite{Scheko04,Iskako07} therefore applies. According to
this phenomenology, $\mathrm{\overline{Rm}}$ can increase as a function of
$\mathrm{\overline{Re}}$ if $\mathrm{Pm}$ is larger than 1 and the eddies near the
viscous length scale generate the magnetic field. If either $\mathrm{Pm}$ is
small or the field generation occurs whithin the inertial range or at the
integral scale, $\mathrm{\overline{Rm}}$ is independent of $\mathrm{\overline{Re}}$. In
spherical Couette flow, the large scales are already capable of dynamo action.
Not surprisingly, the spectra of magnetic energy peak at small $m$ and $l$ (see Fig.
\ref{fig:spectra_compare_B}). Only the computation at the largest $\mathrm{\overline{Re}}$ 
raises doubts because the spectrum of magnetic energy reaches a second local maximum at $l=11$ 
in this case, which nevertheless is still within the inertial range (see Fig.
\ref{fig:spectra_compare_E}) and at any rate below the global maximum at $l=1$.
$\mathrm{Pm}$ is already slightly below 1 in this 
computation so that a further increase in $\mathrm{\overline{Re}}$ should only add a high 
wavenumber tail to the kinetic energy spectrum corresponding to eddies with too small 
a magnetic Reynolds number to contribute to the dynamo effect. In order to further test
whether the small scales are responsible for magnetic field generation, eqs.
(\ref{eq:Induction},\ref{eq:Navier_Stokes}) were solved simultaneously, but the
axisymmetric components of the velocity were removed before the induction term
in eq. (\ref{eq:Induction}) was computed. The magnetic field decayed in this
simulation, which shows that the turbulent eddies in this velocity field are unable
to support the magnetic field by themselves.

Since the magnetic field is generated at large scales even for $\mathrm{Pm}
\approx 1$, an increase in $\mathrm{\overline{Re}}$ and concomittant decrease in
$\mathrm{Pm}$ adds scales which contribute neither to the creation nor the
destruction of the magnetic field. It is concluded that $\mathrm{\overline{Rm}}=600$ is
the critical value of $\mathrm{\overline{Rm}}$ in more turbulent flows as well. In an
experiment with $\mathrm{Pm}=10^{-5}$ this corresponds to $\mathrm{\overline{Re}}=6 \times
10^7$. For Reynolds numbers this large, $E_{\text{kin}}$ is nearly independent of the
Reynolds number according to fig. \ref{fig:blthickness_ns} and equal to 0.023,
from which one deduces $\overline{v_{\text{rms}}}=0.058$ and
$\mathrm{Re} \approx 10^9$. For the sodium experiment in Maryland with
a gapwidth of $d=1m$ and a viscosity
$\nu \approx 10^{-6} m^2/s$, this corresponds to a rotation period of about only
$6 ms$.


To conclude this section, we compare the stability diagram in fig.
\ref{fig:Dynamo_onset_ns} with the results of similar studies. All numerical
simulations report a general increase of the critical $\mathrm{Rm}$ with the
Reynolds number for $\mathrm{Pm}$ larger than 1, an effect which is also found
in the analytical model of ref. \cite{Boldyr04}. The dependence of the dynamo
threshold on the Reynolds number is not monotonous, however, and other published
results have in common with fig. \ref{fig:Dynamo_onset_ns} a maximum in the
critical $\mathrm{Rm}$ for $\mathrm{Pm}$ around 1. This feature is seen in refs.
\cite{Ponty05,Iskako07}. Explanations for this effect have been attempted in refs.
\cite{Brande11,Malysh10} who attributed it to either the bottleneck
effect in turbulent spectra or the helicity at the viscous scale in the velocity field.

Despite the similarities, it is doubtful whether there is a universal mechanism
for the dependence of the dynamo threshold. For example, the plateau in
$\mathrm{Rm}$ is reached for values of $\mathrm{Pm}$ ranging from 0.1 to 1
across the different studies, so that there are variations by an order of
magnitude in the $\mathrm{Pm}$ at which the
critical $\mathrm{Rm}$ is maximum. There is also no indication of a
bottleneck in the spectra presented here which shows that the bottleneck is not
necessary for the observed variation of the dynamo threshold. It remains an open
question why there is a maximum in the critical magnetic Reynolds number as
shown in fig. \ref{fig:Dynamo_onset_ns} in the spherical Couette system.

\section{Rough surface \label{rough}}

\subsection{Hydrodynamic characteristics}

The critical $\mathrm{Re}$ found in the previous section is challenging to
realize in experiments. The problem arises from the fact that the inner sphere
needs to rotate faster in order to increase $\mathrm{\overline{Re}}$. But a faster
rotation rate also thins the boundary layer at the inner sphere and reduces the
volume of fluid directly coupled to the boundary motion. An obvious remedy is to
mount blades on the inner surface. The boundary layer cannot be smaller than the
surface roughness which ensures a better coupling of the fluid to the inner
sphere at high rotation rates. In the following, the blades are modelled by a
volume force as described in section \ref{Math}.

\begin{figure}[h] \centering
\includegraphics[width=0.65\linewidth]{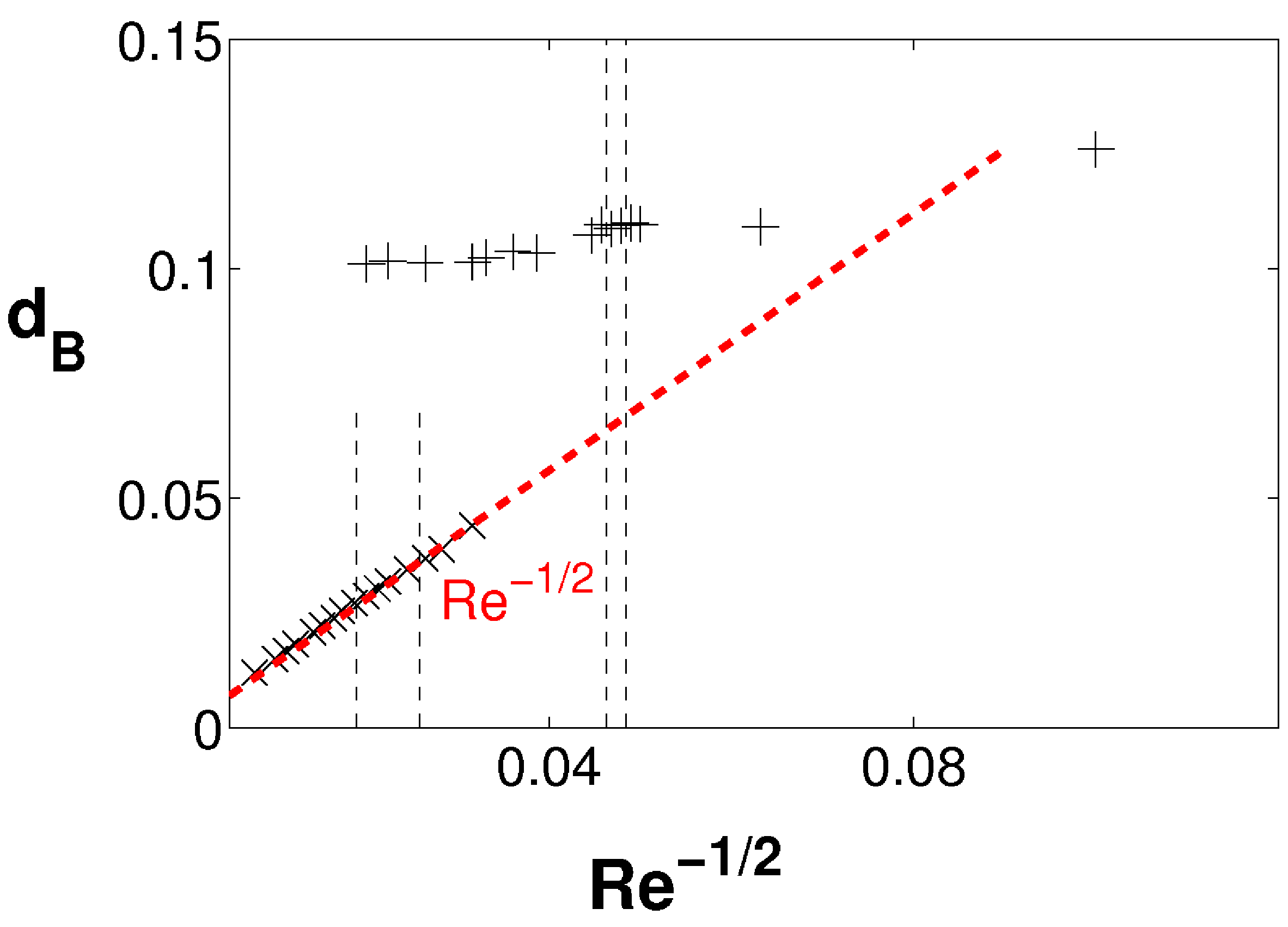}
\hfill
\includegraphics[width=0.65\linewidth]{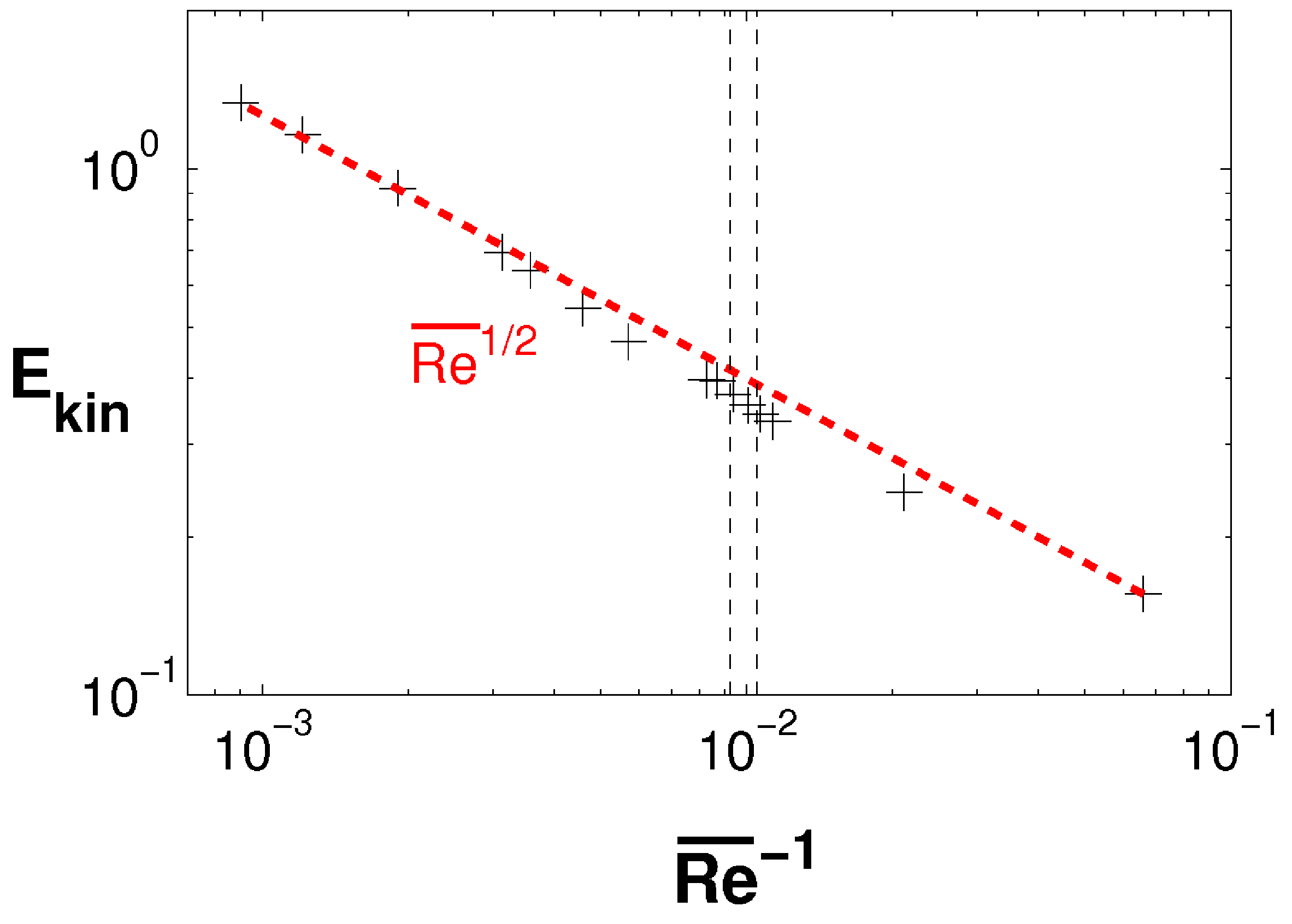} 
\hfill
\includegraphics[width=0.65\linewidth]{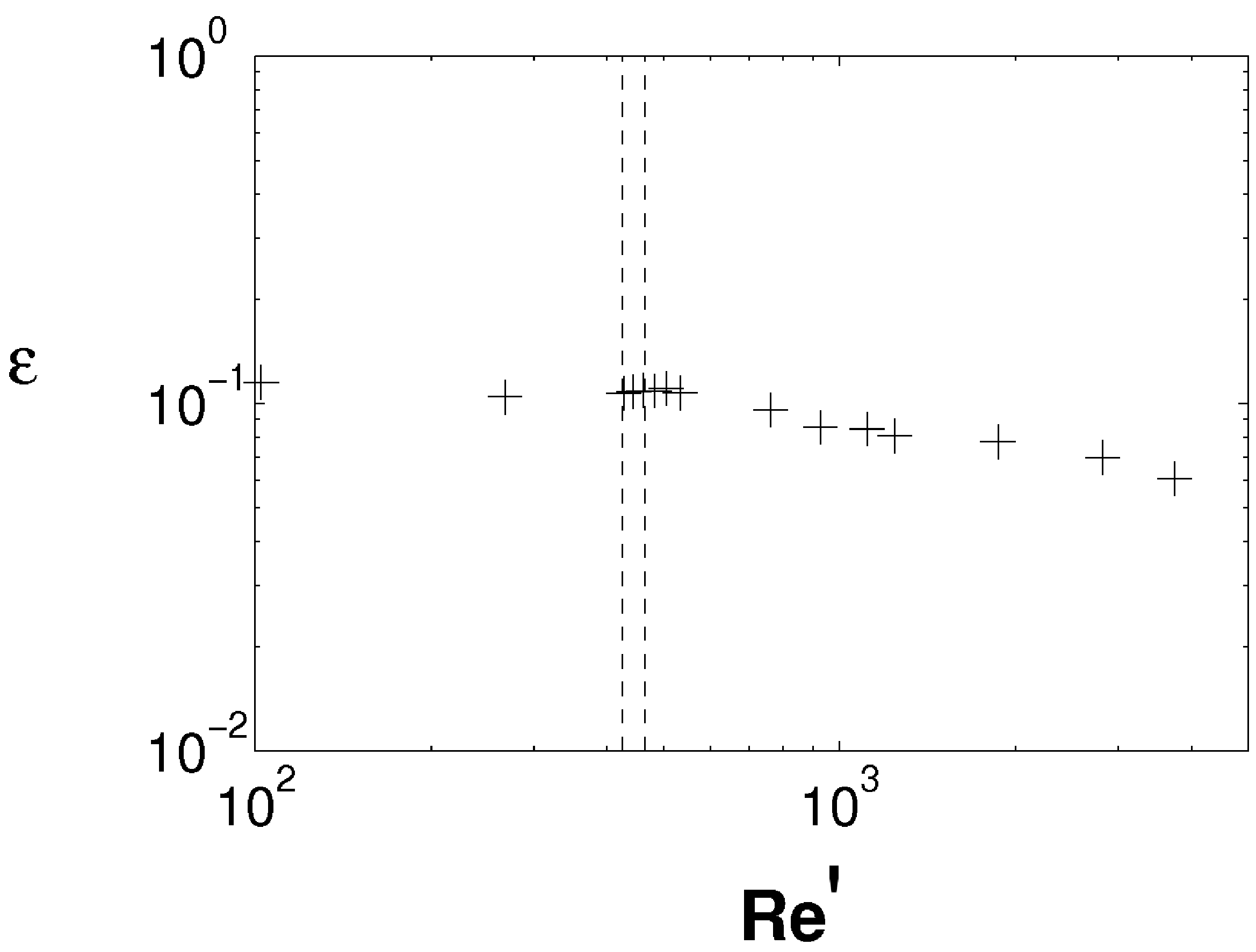} 
\caption{(Color online) Boundary layer thickness $d_B$ plotted against
$\mathrm{Re}^{-1/2}$ (top), kinetic
energy plotted against $\mathrm{\overline{Re}}^{-1}$ (middle), and energy dissipation rate plotted against $\mathrm{Re}'$ (bottom). The vertical dashed lines show transitions
to different regimes. Results obtained for smooth and rough surfaces are
indicated by the symbols $\times$ and +, respectively.}
\label{fig:blthickness_df}
\end{figure}

Due to the increased momentum transport, turbulence develops at lower rotation
rates. The first
nonaxisymmetric instability arises at $\mathrm{Re_h} = 425$
$(\mathrm{\overline{Re_h}} = 95)$ with a dominant wavenumber of $m=2$, and beyond
$\mathrm{Re_s} = 465$ $(\mathrm{\overline{Re_s}} = 108)$ odd wavenumbers increase, too.
The evolution of the boundary layer
thickness is shown in Figure \ref{fig:blthickness_df} (top). As expected,
the thickness of the boundary layer does
not drop below one tenth of the gap width and 
approaches the thickness of the forced layer at large $\mathrm{Re}$. Since the
thickness of the layer in which momentum is injected into the fluid remains
nearly constant, independend from the
rotation rate, the kinetic energy increases with $\mathrm{\overline{Re}}$. It does
so approximately in $\mathrm{\overline{Re}}^{1/2}$, which is shown in Figure
\ref{fig:blthickness_df} (middle).

The torque on the inner sphere is zero for free slip boundary conditions so that
one deduces from the energy budget (\ref{eq:energy_budget}) that the dissipation
rate is given by $\epsilon = \int \bm F \cdot \bm v dV$. As can be seen in the
bottom panel of fig. \ref{fig:blthickness_df}, $\epsilon$ behaves differently
depending on whether the boundary layer is thicker than the forced layer or not.
The interval of Reynolds number in each regime is too small to deduce a 
law relating $\epsilon$ with $\mathrm{Re}'$.

\subsection{Dynamo transition}

\begin{figure}[h] \centering
\includegraphics[width=0.95\linewidth]{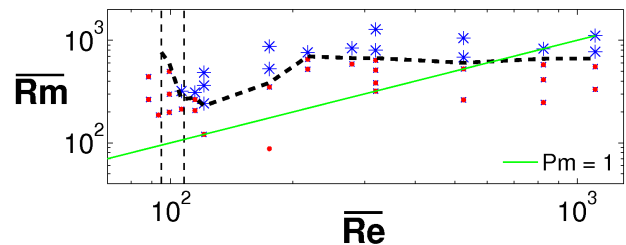} 
\caption{(Color online) Dynamo
transition (black dashed line) with the same symbols as in figure \ref{fig:Dynamo_onset_ns}.}
\label{fig:dynamo_onset_df} \end{figure}

The diagram of the dynamo threshold plotted in the $(\mathrm{\overline{Re}},
\mathrm{\overline{Rm}})-$plane in Figure \ref{fig:dynamo_onset_df} is
similar to Figure \ref{fig:Dynamo_onset_ns}. There is no
dynamo action for axisymmetric flows. Similar to the simulations with
no-slip conditions, even wavenumbers ($m=2$ and harmonics) increase in
the magnetic energy spectra, whereas the total magnetic energy is increasing exponentially  with superposed
oscillations. In contrast to the simulations of the previous section, it was not possible
to find a case in which the odd wavenumbers dominate the magnetic spectrum.
Like in the simulations with no-slip conditions the dynamo
threshold shows a plateau at $\mathrm{\overline{Rm}} = 600$. In experiments with 
$\mathrm{Pm}=10^{-5}$ this corresponds to $\mathrm{\overline{Re}} = 6 \times 10^7$. By extrapolating
the kinetic energy in Figure \ref{fig:blthickness_df} (bottom) up to this $\mathrm{\overline{Re}}$, one gets
$E_{\text{kin}} \approx 300$ and $\overline{v_{\text{rms}}} \approx 6.6$, or $\mathrm{Re}' \approx 9 \times 10^7$ with $\Omega'_i = 9.4$.

\begin{figure}[h] \centering
\includegraphics[width=0.7\linewidth]{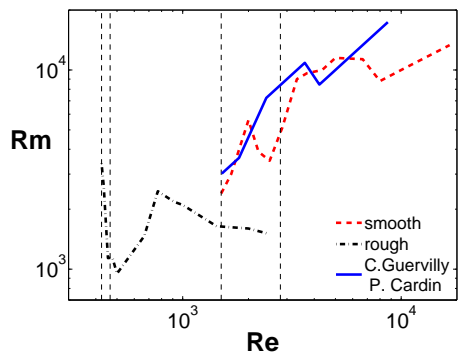}
\caption{(Color online) Dynamo threshold of both surface types and results of \cite{Guervi10}.}
\label{fig:dynamo_comp} \end{figure}

\section{Conclusion}

Laminar spherical Couette flow with a stationary outer sphere is axisymmetric
and consists of a differential rotation superimposed on a meridional circulation,
which flows from the inner to the outer sphere in the equatorial plane,
continues towards the poles of the outer sphere along the outer boundary in each
hemisphere, and finally returns to the inner sphere. Dudley and James
\cite{Dudley89} showed that flows of this type are capable of dynamo action. At
the magnetic Reynolds numbers accessible by numerical computation, laminar
spherical Couette flow nonetheless is not a dynamo due to the unfavourable
velocity profile. For instance, the outward flow is concentrated in a narrow jet
in the equatorial plane. Unstable spherical Couette flow on the other hand
readily generates magnetic field. However, the field generation process still
occurs at large scales, the small turbulent scales on their own do not support
the magnetic field in these dynamos.

The numerical data allow us to extrapolate to more turbulent flows than the
simulated flows if the data are plotted in terms of Reynolds numbers, magnetic
and hydrodynamic, based on the rms velocity rather than the boundary velocity.
The data are more difficult to interpret and extrapolate if they are given as
functions of the Reynolds numbers computed with the inner boundary velocity as
seen in Figure \ref{fig:dynamo_comp}. This plot also includes the data from ref.
\cite{Guervi10}, which should be compared to our simulations with smooth
boundaries. The curves are broadly similar apart from a shift by roughly a
factor 1.5 along the $\mathrm{Re}-$axis, which must be attributed to the different
aspectio ratio and the different magnetic boundary conditions.

If the flow is driven by the volume force which simulates blades of height one
tenth of the gap size mounted on the inner sphere, the dynamo threshold expressed
as critical magnetic Reynolds number based on the rms velocity is reduced by
one sixth. Such a change is plausible because the topology of the flow is the
same as for smooth boundaries, but the kinetic energy is more evenly
distributed in space. Both the boundary layer and the equatorial jet are thicker. From an
experimental point of view, the much more important improvement brought about by
the blades is the reduction of the rotation rate of the inner sphere at the
dynamo onset. The critical Reynolds number based on the rotation frequency of the
inner sphere is reduced by a factor 10 by the blades. 
Using the results obtained with the simulated blades and
the numbers for the Maryland experiment with a gapwidth of one meter
and the viscosity of liquid sodium $\nu \approx 10^{-6} m^2/s$, a Reynolds number number of 
$\mathrm{Re}' = 9 \times 10^7$ 
is reached for a rotation period of the inner sphere of 0.07 $s$.
A spherical Couette dynamo is more readily attainable with blades mounted on the
inner sphere because they improve the coupling between the fluid and the motion
of the inner sphere and at the same time modify little the structure of the
flow.


\end{document}